\documentclass{article}
\usepackage[utf8]{inputenc}
\usepackage[a4paper,top=2cm,bottom=2cm,left=3cm,right=3cm,marginparwidth=1.75cm]{geometry}
\usepackage{gensymb}
\usepackage{graphicx}
\usepackage[colorlinks=true, allcolors=blue]{hyperref}
\usepackage{authblk}
\usepackage{amsmath}
\usepackage{xcolor}
\usepackage[english]{babel}
\usepackage{titlesec}
\usepackage{pdfpages}
\usepackage[section]{placeins}
\titleformat{\paragraph}
{\normalfont\normalsize\bfseries}{\theparagraph}{1em}{}
\titlespacing*{\paragraph}
{0pt}{3.25ex plus 1ex minus .2ex}{1.5ex plus .2ex}
\usepackage[
    backend=biber,
    style=nature,
  ]{biblatex}

\title{Exciton tuning in monolayer WSe$_2$ via substrate induced electron doping}
\author[1,2*]{Yang Pan}
\author[3]{Mahfujur Rahaman}
\author[1,2]{Lu He}
\author[1,2]{Ilya Milekhin}
\author[4]{Gopinath Manoharan}
\author[5]{Muhammad Awais Aslam}
\author[2,4,6]{Thomas Blaudeck}
\author[6]{Andreas Willert}
\author[5]{Aleksandar Matković}
\author[1,2]{Teresa I. Madeira}
\author[1,2]{Dietrich R. T. Zahn}
\affil[1]{Semiconductor Physics, Institute of Physics, Chemnitz University of Technology, Chemnitz, Germany}
\affil[2]{Center for Materials, Architectures, and Integration of Nanomembranes (MAIN), Chemnitz University of Technology, Chemnitz, Germany}
\affil[3]{Department of Electrical and Systems Engineering, University of Pennsylvania, Philadelphia, PA, USA}
\affil[4]{Center for Microtechnologies, Chemnitz University of Technology, Chemnitz, Germany}
\affil[5]{Institute of Physics, Montanuniversität Leoben, Leoben, Austria}
\affil[6]{Fraunhofer Insitute for Electronic Nano Systems, Chemnitz, Germany}
\affil[*]{Corresponding author: yang.pan@physik.tu-chemnitz.de}

\date{}

\begin{document}

\maketitle

\begin{abstract}
We report on large exciton tuning in WSe$_2$ monolayers via substrate induced non-degenerate doping. We observe a redshift of $\sim$62 meV for the $A$ exciton together with a 1-2 orders of magnitude photoluminescence (PL) quenching when the monolayer WSe$_2$ is brought in contact with highly oriented pyrolytic graphite (HOPG) compared to the dielectric substrates such as hBN and SiO$_2$. As the evidence of doping from HOPG to WSe$_2$, a drastic increase of the trion emission intensity was observed. Using a systematic PL and Kelvin probe force microscopy (KPFM) investigation on WSe$_2$/HOPG, WSe$_2$/hBN, and WSe$_2$/graphene, we conclude that this unique excitonic behavior is induced by electron doping from the substrate. Our results propose a simple yet efficient way for exciton tuning in monolayer WSe$_2$, which plays a central role in the fundamental understanding and further device development.
    
\end{abstract}

\section{Introduction}
Beyond graphene \cite{novoselov2004electric}, transition metal dichalcogenides (TMDCs) are currently at the center of 2D materials research, owing to their extraordinary optical, electrical, thermal, mechanical properties \cite{wilson1969transition,wu2013vapor,peimyoo2015thermal,li2017enhancement}, and, most importantly, to the unique indirect- to direct-bandgap transition when the material is thinned from bulk to monolayer \cite{splendiani2010emerging,li2018accurate}. Different approaches of exciton tuning and bandgap engineering have been reported such as changing the dielectric environment, mechanical straining, doping, alloying, injecting plasmonic hot electrons, and manipulating the carrier concentration via external electric field \cite{raja2017coulomb,gong2014band,conley2013bandgap,frisenda2017biaxial,li2015active,nguyen2019visualizing,fernandez2020renormalization,chaves2020bandgap}. 

 In this work, we report on tuning the exciton energy in monolayer WSe$_2$ via substrate induced non-degenerate electron doping. We observe a $\sim$62 meV redshift of the monolayer WSe$_2$ $A$ excitonic emission (from $\sim$1.65 eV to $\sim$1.71 eV) together with a few orders of magnitude photoluminescence (PL) quenching when the material is brought in contact with HOPG compared to the WSe$_2$ excitonic feature on dielectric substrates such as hBN, SiO$_2$, and polydimethylsiloxane (PDMS), which has been measured and reported in our previous work \cite{tonndorf2013photoluminescence}. As a by-product, a drastic increase of the trion emission intensity up to 5.5 times was observed, which is a characteristic of electron doping in TMDC monolayers. To understand this unique behavior, we employed a systematic PL and Kelvin probe force microscopy (KPFM) investigation on different TMDC/substrate combinations, namely WSe$_2$/HOPG, WSe$_2$/graphene, and WSe$_2$/hBN as a reference. Surprisingly, we were only able to observe such pronounced redshift when WSe$_2$ is in contact with HOPG but not with graphene. The KPFM measurements provide different contact potential difference (CPD) values when comparing WSe$_2$/HOPG to WSe$_2$/graphene and WSe$_2$/hBN. This indicates different Fermi level positions and different carrier concentrations in WSe$_2$. The PL quenching, redshift, increase of trion emission intensity, and different CPD values all conclusively point towards electrons from the HOPG substrate injected to WSe$_2$ and leading to bandgap renormalization and thus the tuning of exciton energy. Our work explains the unique behavior of monolayer WSe$_2$/HOPG and demonstrates a simple yet efficient method, which enables to tune the exciton energy in monolayer WSe$_2$ by $\sim$62 meV. This is essential for fundamental studies and the development of devices such as photodetectors, excitonic LEDs, and the coupling with plasmonics \cite{zhou2020self,xiao2017excitons,liu2016strong,ross2014electrically}.

\section{Materials and methods}
\subsection{Sample preparation}
Few layer hBN (from 2D semiconductors), graphene (from NGS Naturgraphit), and monolayer WSe$_2$ (from HQ graphene) are mechanically exfoliated from their bulk crystals via Nitto tape onto a PDMS stamp and then transferred bottom-to-top onto the HOPG substrate following a deterministic all-dry transfer technique \cite{castellanos2014deterministic,kunstmann2018momentum}. All materials on PDMS are first characterized by PL and Raman prior to transfer. HOPG was cleaved before transfer to ensure a fresh surface. After transfer, the samples are annealed in a nitrogen atmosphere at 150 \degree C for 2 hours to optimize the contact between flakes and ensure a clean surface. The detailed process used for sample fabrication is shown in Figs. 1S and 2S.

\subsection{Optical spectroscopy}
PL measurements are performed using a Horiba Xplora Plus equipped with a 100x, 0.9 NA objective, a spectrometer comprising 600 l/mm grating, and an electron-multiplying CCD (EMCCD). A DPSS 532 nm CW laser source was used for excitation. The laser power is $\sim$100 µW measured under the objective for PL measurements if not specified differently. The setup is equipped with a Märzhäuser motorized $xyz$ stage with a 100 nm step size precision for PL mapping.

Raman spectra are acquired by a Horiba LabRAM HR spectrometer with a 100x, 0.9 NA objective, 2400 l/mm grating, and a liquid nitrogen cooled Symphony CCD detector. A solid-state 514.7 nm laser is used for excitation with a laser power $\sim$100 µW measured under the objective. We choose a confocal pinhole of 50 µm to reach a high spectral resolution of approximately $0.8~ cm^{-1}$.

\subsection{Kelvin probe force microscope}
We use an AIST-NT SmartSPM$^{TM}$ 1000 for KPFM measurements. The KPFM measurements are performed in ambient condition with constant temperature and humidity. The NSG10 Pt coated tip is commercially available with a typical tip radius of $\sim$35 nm.

\section{Results and discussion}

\begin{figure}
    \centering
    \includegraphics[width=1\textwidth]{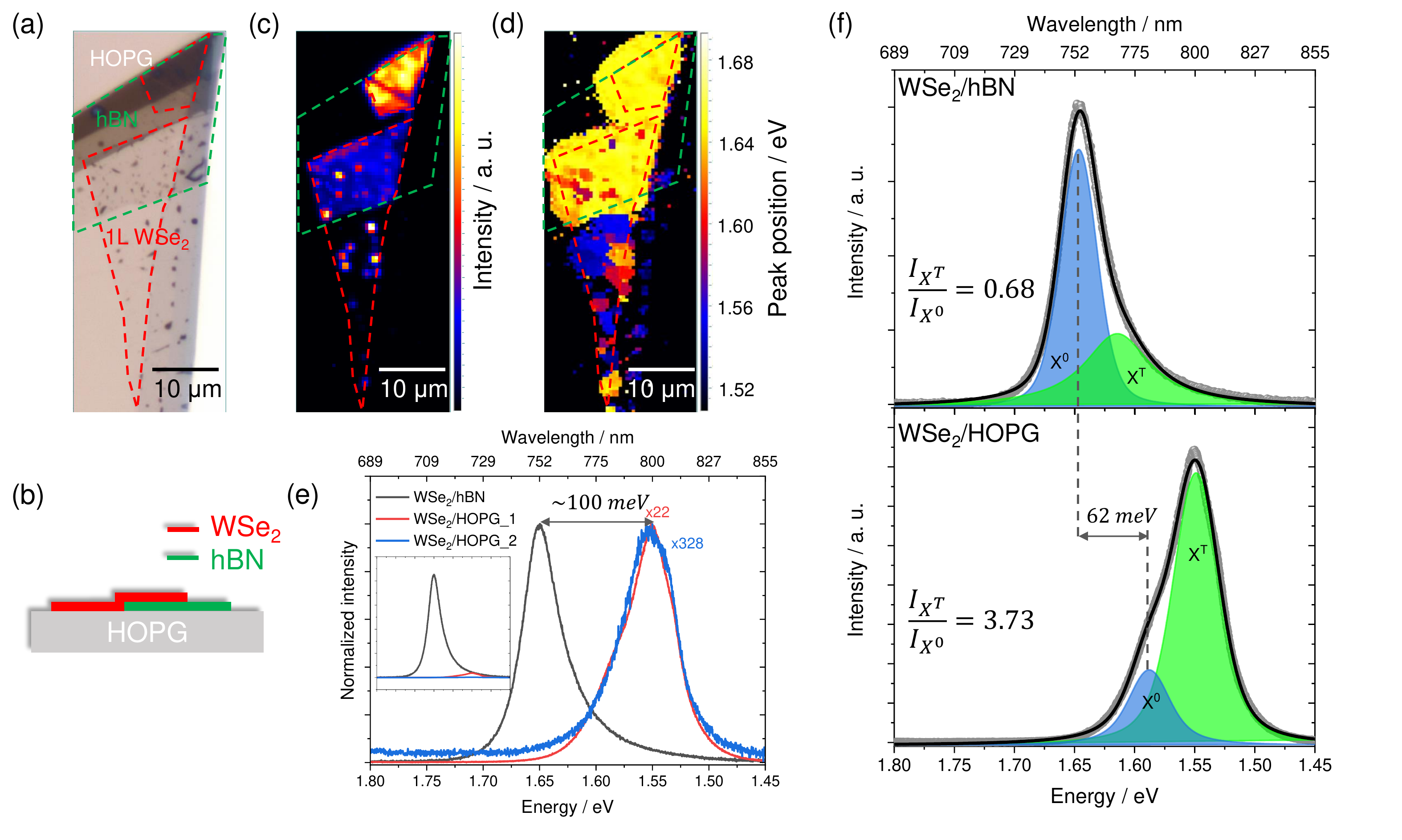}
    \caption{(a) optical microscope image and (b) schematic cross section of the WSe$_2$/hBN/HOPG hetero-stack. PL (c) intensity and (d) peak position map of the sample. (e) PL spectra of WSe$_2$/hBN and WSe$_2$/HOPG. For comparison, the PL intensity of WSe$_2$/HOPG is normalized to that of WSe$_2$/hBN. Inset: as-measured (not-normalized) PL spectra. (f) Fitted PL spectra of WSe$_2$/hBN and WSe$_2$/HOPG.}
    \label{fig1}
\end{figure}

Figs. \ref{fig1}(a) and (b) display the optical microscope image and the schematic cross section of a WSe$_2$/hBN/HOPG hetero-stack, respectively. The monolayer WSe$_2$ is transferred onto the hBN/HOPG hetero-stack in a way that it creates contacts with both few layer hBN and HOPG. According to the atomic force microscopy (AFM) measurement shown in Fig. 4S, the top brown-colored hBN has a thickness of $\sim$38.2 nm, and the middle part has a thickness of $\sim$3.8 nm. We acquired a micro PL map on the sample with a step size of 0.5 µm. As shown in the intensity map in Fig. \ref{fig1}(c), the PL intensity of WSe$_2$ on thick hBN is higher than that on thin hBN because of interference enhancement \cite{ding2018understanding,rojas2021photoluminescence}. More importantly, comparing the PL intensity of WSe$_2$ on hBN and HOPG, one can clearly identify that a drastic decrease of PL intensity occurs on HOPG. The few dots that still remain intense may correspond to bubbles or hydrocarbon contaminations at the interface, which can enhance the PL signal \cite{purdie2018cleaning,haigh2012cross,tyurnina2019strained}. Fig. \ref{fig1}(d) displays the peak position map indicating that the sample is clearly divided into two parts: WSe$_2$/hBN with a peak position of $\sim$1.65 eV and WSe$_2$/HOPG with a peak position of $\sim$1.55 eV.

The detailed spectra of WSe$_2$/hBN and WSe$_2$/HOPG are shown in Fig. \ref{fig1}(e). A strong PL quenching of 1-2 orders of magnitude is observed when WSe$_2$ is in contact with HOPG, which indicates charge dissociation through the junction or charge transfer between WSe$_2$ and HOPG \cite{rojas2021photoluminescence,hwang2021interlayer}. Monolayer WSe$_2$ on hBN shows a characteristic PL at $\sim$1.65 eV, which is consistent with the literature values \cite{tonndorf2013photoluminescence,huang2016probing}, while the PL peak position of WSe$_2$/HOPG shows a marked $\sim$100 meV redshift, which is much higher than the reported value caused by changing of dielectric environment \cite{raja2017coulomb}. Besides the quenching and redshift, the PL line shape changes significantly. We thus decomposed the PL spectra into peaks corresponding to the radiative recombination of different exciton/trion species in monolayer WSe$_2$. As shown in Fig. \ref{fig1}(f), two peaks with a Voigt line shape were identified in the fitted spectra. The neutral exciton ($X^0$) originates from the direct bandgap transition at the $K$ point in the Brillouin zone and there is a charged exciton peak also known as trion $X^T$ \cite{huang2016probing,liu2019valley,he2020valley,li2020fine}. We also investigated the Stokes shift of monolayer WSe$_2$ as shown in Fig. 3S, which is negligible with value of $\sim$2 meV. It is therefore fair enough to consider the PL peak position as the exciton energy. The fitting result suggests a 62 meV redshift of $X^0$ and most interestingly, a drastic increase of the relative $X^T$ intensity. The ratio of $I_{X^T}/I_{X^0}$ increases from 0.68 on hBN to 3.73 on HOPG, which is a strong evidence of higher electron concentration in WSe$_2$ on HOPG than in WSe$_2$ on hBN.

\begin{figure}
    \centering
    \includegraphics[width=1\textwidth]{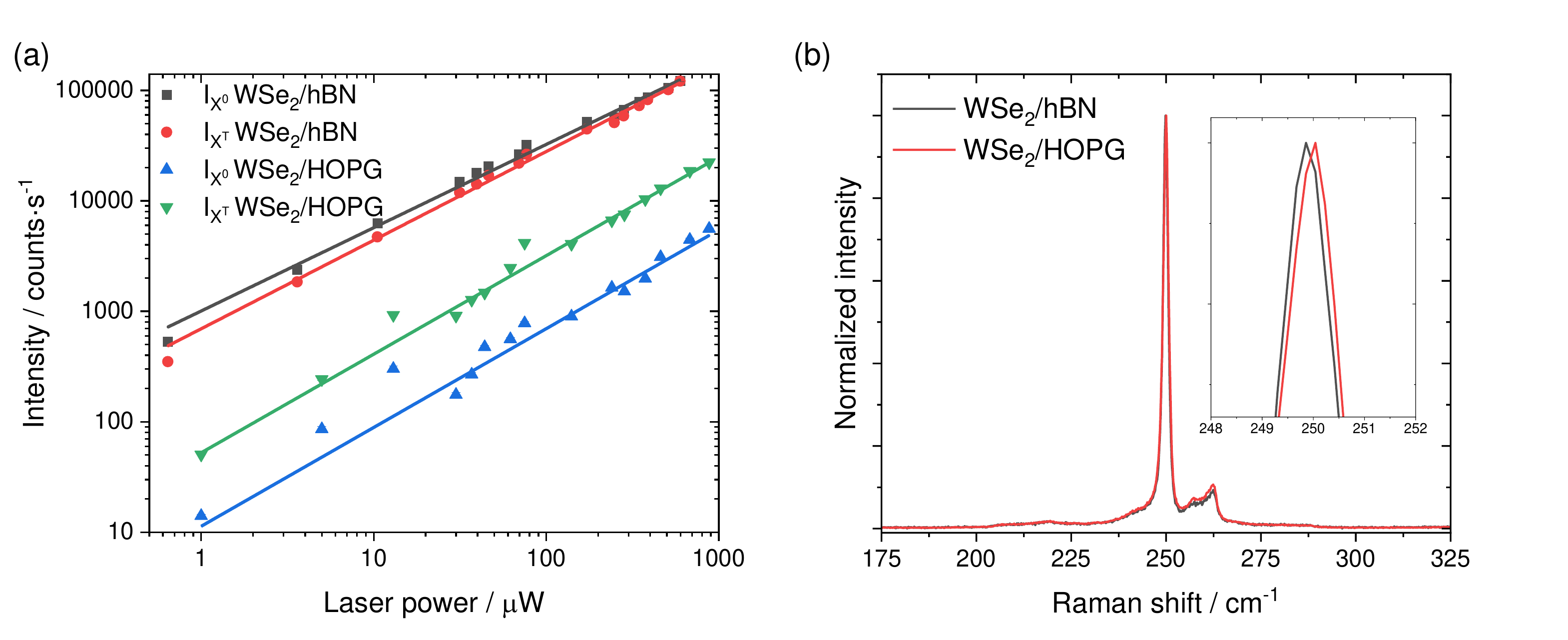}
    \caption{(a) PL intensity as a function of excitation power for $X^0$ and $X^T$ emissions from WSe$_2$/hBN and WSe$_2$/HOPG. Solid lines are fits to a power law: $I\propto P^{\alpha}$. (b) High spectral resolution Raman spectra of WSe$_2$/hBN and WSe$_2$/HOPG. Inset is a zoom in at 248-252 $cm^{-1}$.}
    \label{fig2}
\end{figure}

Even though we propose that charge transfer and electron doping from HOPG to monolayer WSe$_2$ seem to be the most reasonable mechanism of PL quenching, redshift, and increasing trion emission intensity, we still carefully examined that they do not originate from the defect-bound localized states or strain due to lattice mismatch. Power dependent PL intensities of WSe$_2$/hBN and WSe$_2$/HOPG are displayed in Fig. \ref{fig2}(a). The PL intensity is obtained from the integrated area of the Voigt fitted $X^0$ and $X^T$ features. The PL intensity as a function of excitation laser power is then fitted by a power law: $I\propto P^{\alpha}$  \cite{huang2016probing,wu2016defects}, where the extracted exponential factor $\alpha_{{X^0}_{WSe_2/hBN}}=0.75\pm0.02$, $\alpha_{{X^T}_{WSe_2/hBN}}=0.80\pm0.01$, $\alpha_{{X^0}_{WSe_2/HOPG}}=0.89\pm0.03$, and $\alpha_{{X^T}_{WSe_2/HOPG}}=0.89\pm0.05$ for $X^0$ and $X^T$ on WSe$_2$/hBN and WSe$_2$/HOPG, respectively. The fitting results suggest a sublinear power dependence of the PL intensity for both $X^0$ and $X^T$ on WSe$_2$/hBN and WSe$_2$/HOPG and do not show any saturation phenomena at high laser power, which excludes the possibility of defects as the origin of the observed behavior \cite{wu2017spectroscopic}. Fig. \ref{fig2}(b) shows the high spectral resolution ($\sim$0.8 $cm^{-1}$) Raman spectra of WSe$_2$/hBN and WSe$_2$/HOPG. The most intense peak at $\sim$250 $cm^{-1}$ corresponds to the combination of the in-plane $E_{2g}$ and out-of-plane $A_{1g}$ vibrational modes, which are almost degenerate at the same frequency \cite{tonndorf2013photoluminescence,luo2013effects,zhao2013lattice,terrones2014new}. The feature at $\sim$260 $cm^{-1}$ is a second order peak caused by a double resonance effect involving the longitudinal acoustic phonon at the $M$ point in the Brillouin zone assigned as 2LA(M) \cite{terrones2014new,del2014excited}. The $E_{2g}/A_{1g}$ mode is highly sensitive to the strain \cite{dadgar2018strain,desai2014strain}. The fitted Raman spectra reveal a small 0.15 $cm^{-1}$ peak position difference, which indicates that strain is also small and cannot account for the huge redshift in PL.


KPFM is a powerful technique to obtain local surface potential and Fermi level position in the nanoscale \cite{rojas2021photoluminescence,melitz2011kelvin}. We therefore measured KPFM on the WSe$_2$/hBN/HOPG hetero-stack to obtain further insight in the energy level alignment at the various interfaces. In the ideal case KPFM measures the contact potential difference (CPD) between the metallic AFM tip and the sample according to the relation: $CPD=(\phi_{sample}-\phi_{tip})/e$, where $\phi_{sample}$ and $\phi_{tip}$ are the work functions of the sample and the tip, and $e$ is the elementary charge. KPFM does not give a quantitative, absolute value of the Fermi level position in ambient conditions, because the CPD value is known to be strongly influenced by the measurement environment, tip geometry, parasitic effects such as capacitive coupling, as well as the chosen experimental parameters \cite{giusca2015water,jacobs1998resolution,barbet2014cross,jacobs1999practical}. Nevertheless, it still qualitatively indicates the trend of Fermi level position and material work functions \cite{rojas2021photoluminescence,melitz2011kelvin,castanon2020calibrated,matkovic2020interfacial}. The values of the energy levels discussed in the following paragraph are directly extracted from the KPFM measurements. 

\begin{figure}[!hbt]
    \centering
    \includegraphics[width=1\textwidth]{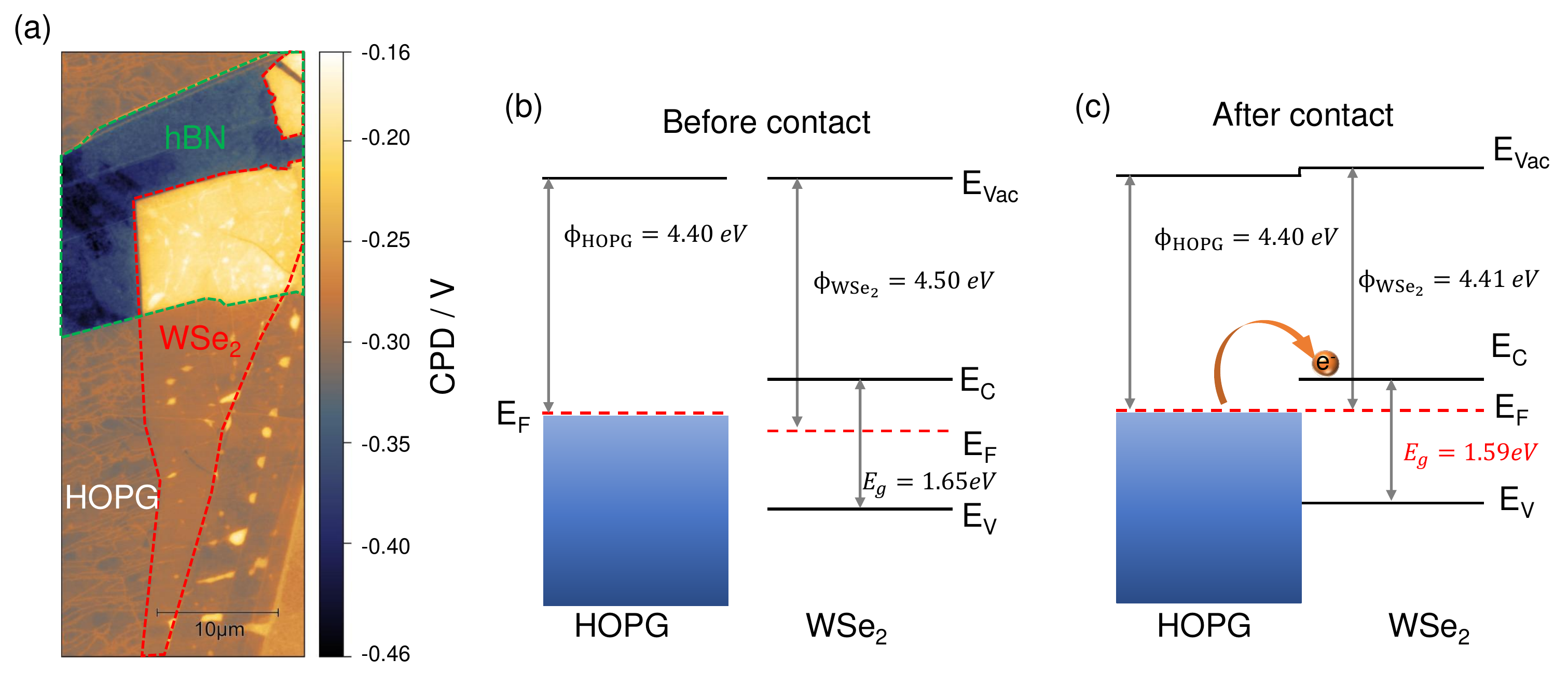}
    \caption{(a) KPFM of WSe$_2$/hBN/HOPG. Band diagram of monolayer WSe$_2$ and HOPG before (b) and after (c) contact. Before contact means when WSe$_2$ is isolated from HOPG by hBN and after contact means that WSe$_2$ is on HOPG.}
    \label{fig3}
\end{figure}

Fig. \ref{fig3}(a) shows the CPD map of WSe$_2$/hBN/HOPG. Even though it is the same monolayer WSe$_2$ flake, one can clearly distinguish the high contrast between WSe$_2$ on hBN and WSe$_2$ on HOPG. The absolute work function of HOPG is determine to be $\phi_{HOPG}=4.4~eV$ by an ultraviolet photoelectron spectroscopy (UPS) measurement shown in Fig. 6S. The electron affinity of monolayer WSe$_2$ is reported to be 3.7-3.9 eV \cite{xiao2018enhanced,liu2013high}. We therefore calculate and draw the band diagrams of WSe$_2$ before (on hBN) and after (on HOPG) contacting with HOPG in Figs. \ref{fig3}(b) and (c), respectively. The band diagrams reveal a decrease of the work function or increase of Fermi level when WSe$_2$ is in contact with HOPG, which indicates higher electron concentration in WSe$_2$ on HOPG than in WSe$_2$ on hBN. The high electron concentration in WSe$_2$/HOPG can only originate from electron doping from HOPG to WSe$_2$, which explains the PL quenching, redshift, and increasing trion emission intensity shown in Fig. \ref{fig1}.

\begin{figure}
    \centering
    \includegraphics[width=1\textwidth]{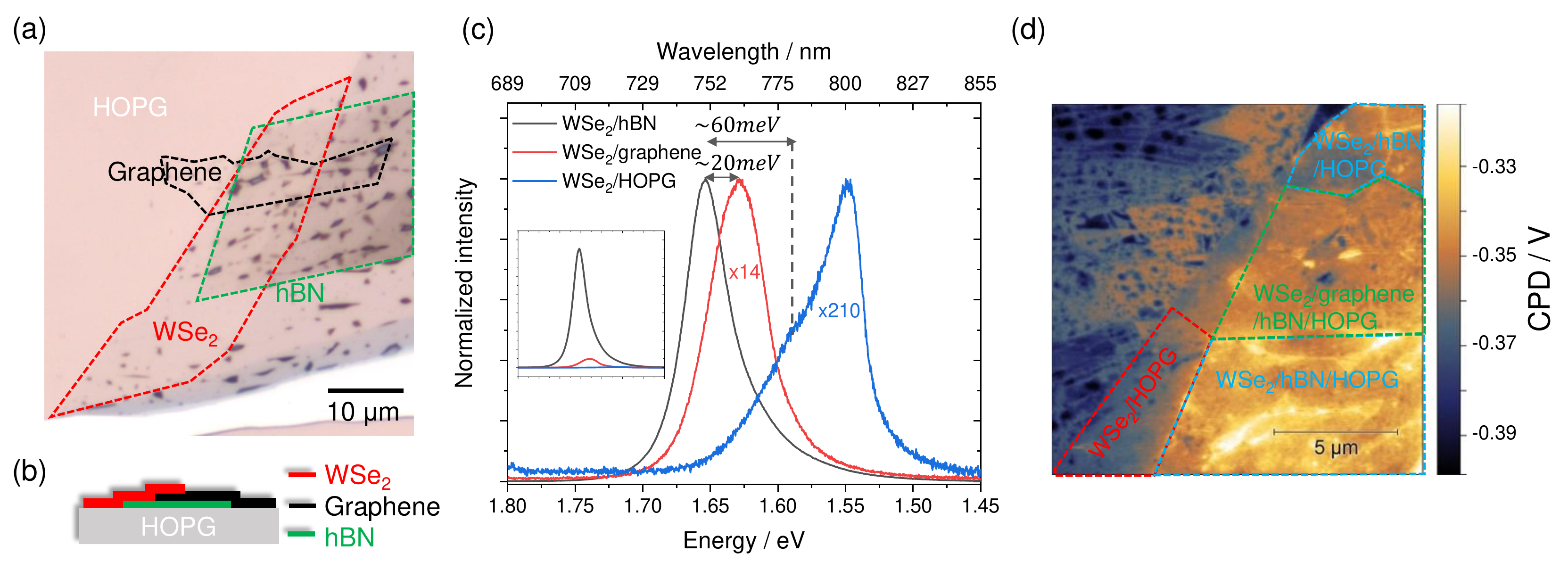}
    \caption{(a) optical microscope image and (b) schematic cross section of WSe$_2$/graphene/hBN/HOPG hetero-stack. (c) PL spectra of WSe$_2$/hBN, WSe$_2$/graphene, and WSe$_2$/HOPG. For comparison, the intensities of the WSe$_2$/graphene and WSe$_2$/HOPG PL are normalized to that of WSe$_2$/hBN/HOPG. Inset: as-measured (not-normalized) PL spectra.  (d) KPFM of WSe$_2$/graphene/hBN/HOPG.}
    \label{fig4}
\end{figure}

Apparently interfacing WSe$_2$ with HOPG results in an efficient tuning of the exciton emission in a straightforward manner. Researchers also studied the combination of WSe$_2$ and graphene \cite{raja2017coulomb}, yet did not report similar results. This naturally leads to the question: do graphene and graphite lead to a different interaction when interfaced with WSe$_2$? To answer this question, we prepared a hetero-stack of WSe$_2$/graphene/hBN/HOPG as shown in Figs. \ref{fig4}(a) and (b), where WSe$_2$ is partially on hBN, partially on graphene, and partially on HOPG. The PL spectra of WSe$_2$/hBN, WSe$_2$/graphene, and WSe$_2$/HOPG are shown in Fig. \ref{fig4}(c). Again, we observe similar PL quenching, redshift, and increasing trion emission intensity for WSe$_2$ on HOPG. However, a redshift of only 20 meV is detected on WSe$_2$/graphene, which is in excellent agreement with the value reported by Raja \textit{et al.} \cite{raja2017coulomb}. This redshift of the $A$ exciton is attributed to the altered local dielectric screening of the Coulomb interaction in WSe$_2$. A higher trion emission intensity is also not observed in the case of WSe$_2$/graphene. This clearly indicates that the interaction for WSe$_2$ on graphene is different from that for WSe$_2$ on HOPG. We assume that this difference is due to the lower amount of free electrons in graphene than that in HOPG. The KPFM measurement performed on such a sample is displayed in Fig. \ref{fig4}(d). A CPD contrast is only observed between WSe$_2$/HOPG and WSe$_2$/hBN with a value of $\Delta CPD_{WSe_2/hBN-WSe_2/HOPG}=(20.4\pm4.9)~mV$, while WSe$_2$/graphene and WSe$_2$/hBN reveal a negligible difference of $\Delta CPD_{WSe_2/hBN-WSe_2/gr}=(2.4\pm4.3)~mV$. This suggests that a significant change of the Fermi level position occurs due to electron doping from the substrate and only happens for WSe$_2$ on HOPG but not for WSe$_2$ on graphene.

\section{Conclusions}
In summary, we investigated WSe$_2$/hBN, WSe$_2$/graphene, and WSe$_2$/HOPG hetero-stacks. We observed a strong PL intensity quenching, 62 meV redshift of the $A$ exciton, and a drastic increase of the trion emission intensity on WSe$_2$/HOPG compared to WSe$_2$/graphene and WSe$_2$/hBN. The KPFM results reveal a high CPD contrast, which indicates a renormalization of the energy level alignment at the interface. The effects observed for WSe$_2$ on HOPG are thus assigned to significant electron doping of the WSe$_2$ monolayer from the HOPG substrate. We propose a simple yet efficient way to tune the exciton emission in monolayer WSe$_2$ by substrate induced electron doping.

\section{Acknowledgments}
The authors gratefully acknowledge financial support by the Deutsche Forschungsgemeinschaft (DFG, projects ZA 146/43-1 and ZA 146/47-1). M.A.A. and A.M. acknowledge financial support by the Austrian Science Fund (FWF) under the grant no. I4323-N36. We thank Manuel Monecke for helping the UPS measurement.

\section{Author contributions}
Y.P. fabricated the samples, performed the measurements and analyzed the data. M.R., I.M. L.H., and T.I.M. contributed to data analysis and discussion. G.M., T.B. and A.W. performed the reflectance contrast measurement. M.A.A. and A.M. provided the graphene.
D.R.T.Z. supervised the work. M.R. and D.R.T.Z. were involved in the evaluation and interpretation of the results. Y.P. wrote the manuscript. All authors discussed the results and commented on the manuscript.

\section{Conflict of interest}
The authors declare no conflict of interest.

\printbibliography


\section*{Supplementary material}

# Supplementary information: Exciton tuning in monolayer WSe$_2$ via substrate induced electron doping

Yang Pan[1,2*], Mahfujur Rahaman[3], Lu He[1,2], Ilya Milekhin[1,2], Gopinath Manoharan[4], Muhammad Awais Aslam[5], Thomas Blaudeck[2,4,6], Andreas Willert[6], Aleksandar Matković[5], Teresa I. Madeira[1,2], and Dietrich R. T. Zahn[1,2]

[1]Semiconductor Physics, Institute of Physics, Chemnitz University of Technology, Chemnitz, Germany
[2]Center for Materials, Architectures, and Integration of Nanomembranes (MAIN), Chemnitz University of Technology, Chemnitz, Germany
[3]Department of Electrical and Systems Engineering, University of Pennsylvania, Philadelphia, PA, USA
[4]Center for Microtechnologies, Chemnitz University of Technology, Chemnitz, Germany
[5]Institute of Physics, Montanuniversität Leoben, Leoben, Austria
[6]Fraunhofer Insitute for Electronic Nano Systems, Chemnitz, Germany
*Corresponding author: yang.pan@physik.tu-chemnitz.de

# 1 Sample preparation

## 1.1 WSe$_2$/hBN/HOPG hetero-stack

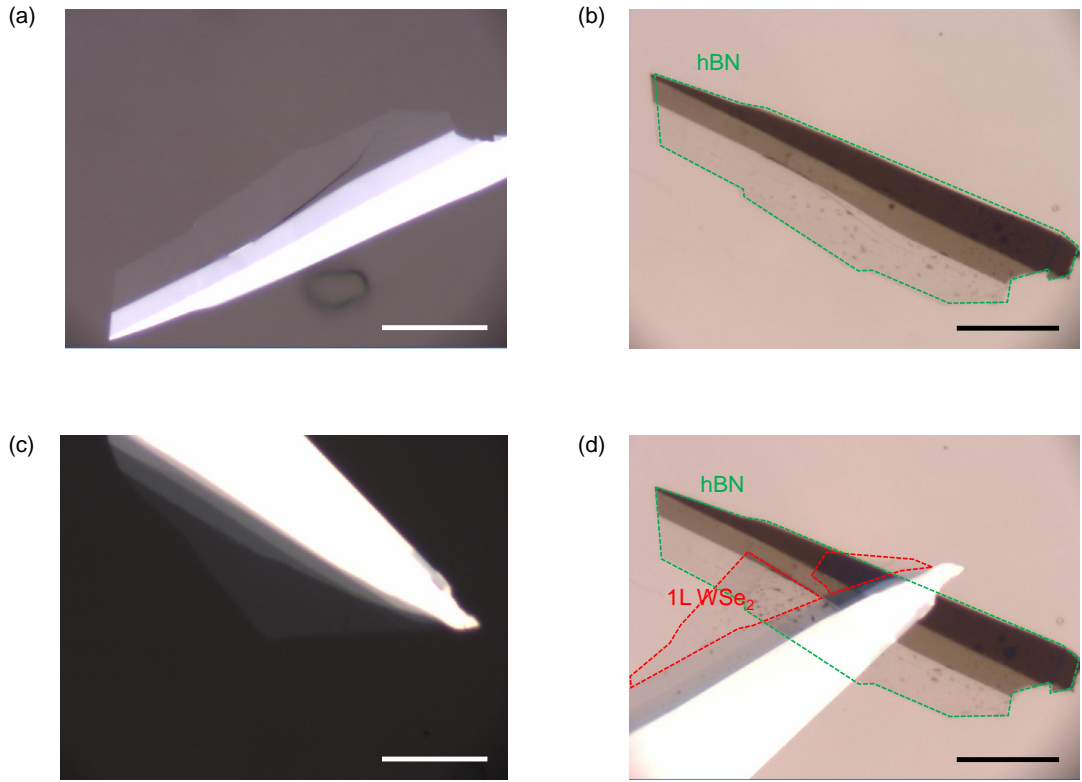

Figure 1S: Optical microscope images of few layer hBN on (a) PDMS, (b) HOPG and monolayer WSe$_2$ on (c) PDMS, (d) hBN/HOPG. Scale bar in figure is 20 µm.

Monolayer WSe$_2$ and few layer hBN are mechanical exfoliated from their bulk materials via Nitto tape on PDMS stamp (as shown in Fig. 1S (a) and (c)). WSe$_2$ is firstly characterized by PL and Raman spectroscopy to identify the layer numbers before transfer. After confirming the layer numbers, the HOPG top layer is cleaved to ensure a clean surface. The hBN and WSe$_2$ are immediately transferred bottom-to-top with a all-dry deterministic transfer technique[1].

## 1.2 WSe$_2$/graphene/hBN/HOPG hetero-stack

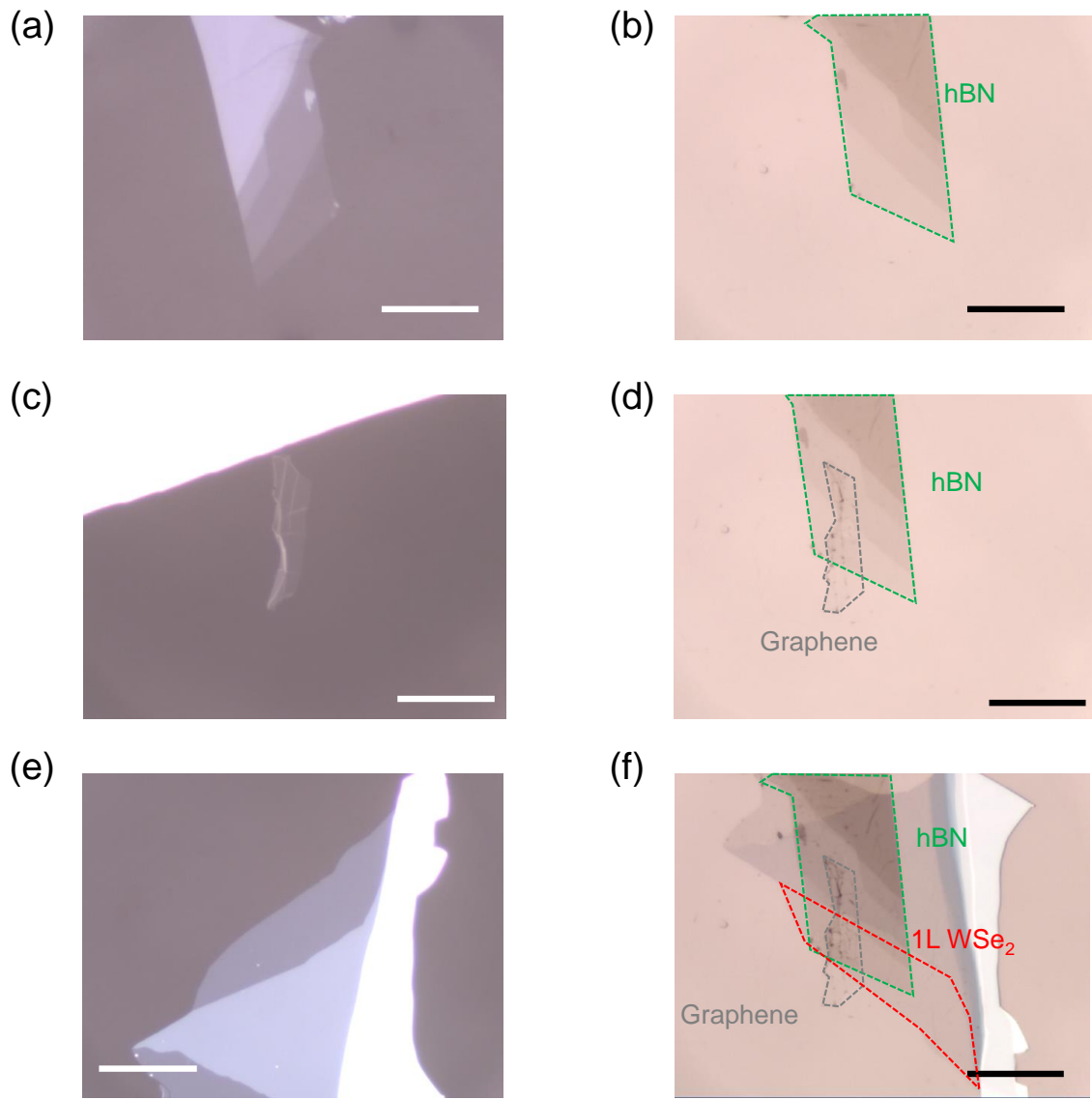

Figure 2S: Optical microscope images of few layer hBN on (a) PDMS, (b) HOPG, graphene on (c) PDMS, (d) hBN/HOPG and monolayer WSe$_2$ on (e) PDMS, (f) graphene/hBN/HOPG. Scale bar in figure is 20 µm.

The sample preparation procedure is same as mentioned above.

## 2 Stokes shift of monolayer WSe$_2$

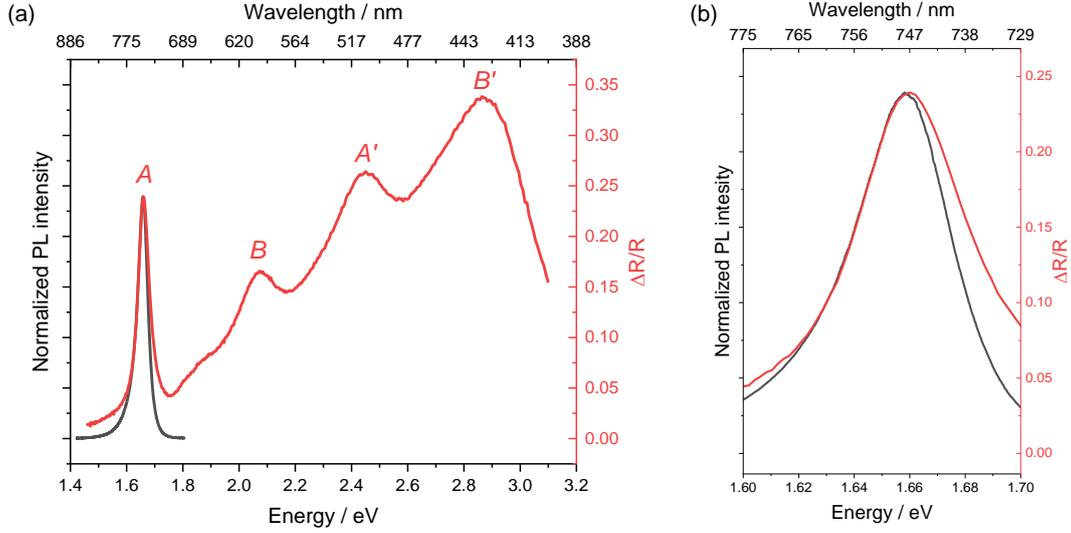

Figure 3S: (a) Micro PL and micro reflectance contrast spectra of monolayer WSe$_2$. (b) zoomed in for 1.60-1.70 eV.

Micro reflectance contrast measurements are carried out with a Zeiss AxioImager.M2m microscope in epi-illumination configuration equipped with a 50x, 0.75NA objective, a Zeiss HAL 100 illuminator-12 V/ 100W white-light source with intensity control and coupled to a J&M Analytik AG Tidas S MSP 800 spectrometer operable in the spectral range 200-980 nm [2, 3].

For the ultra-thin film on a transparent substrates, $\Delta R/R$ is predominantly determined by the imaginary part of the dielectric function, which is proportional to the optical absorption[4–7].

We measured the micro PL and micro reflectance contrast spectra to extract the Stokes shift of monolayer WSe$_2$ to make sure that it is reasonable to consider the PL peak energy position corresponding to the exciton energy. As shown in Fig. 3S, we only observe a ∼2 meV Stokes shift, which makes it fair enough to consider the exciton PL peak position as the exciton energy.

# 3 AFM and KPFM

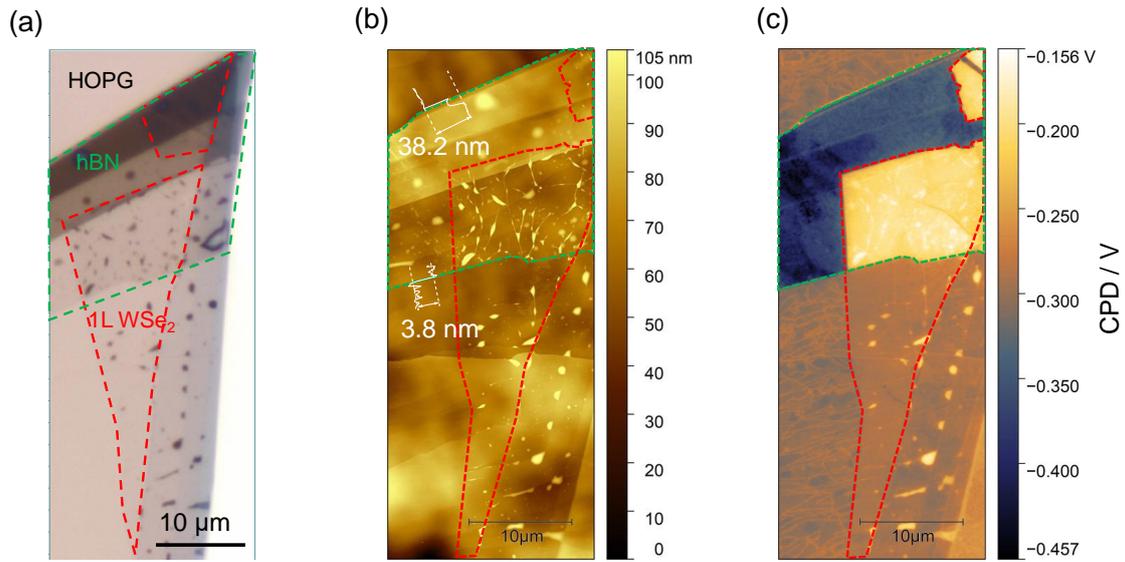

Figure 4S: (a) optical microscope image, (b) AFM height image, and (c) KPFM image of $WSe_2$/hBN/HOPG hetero-stack.

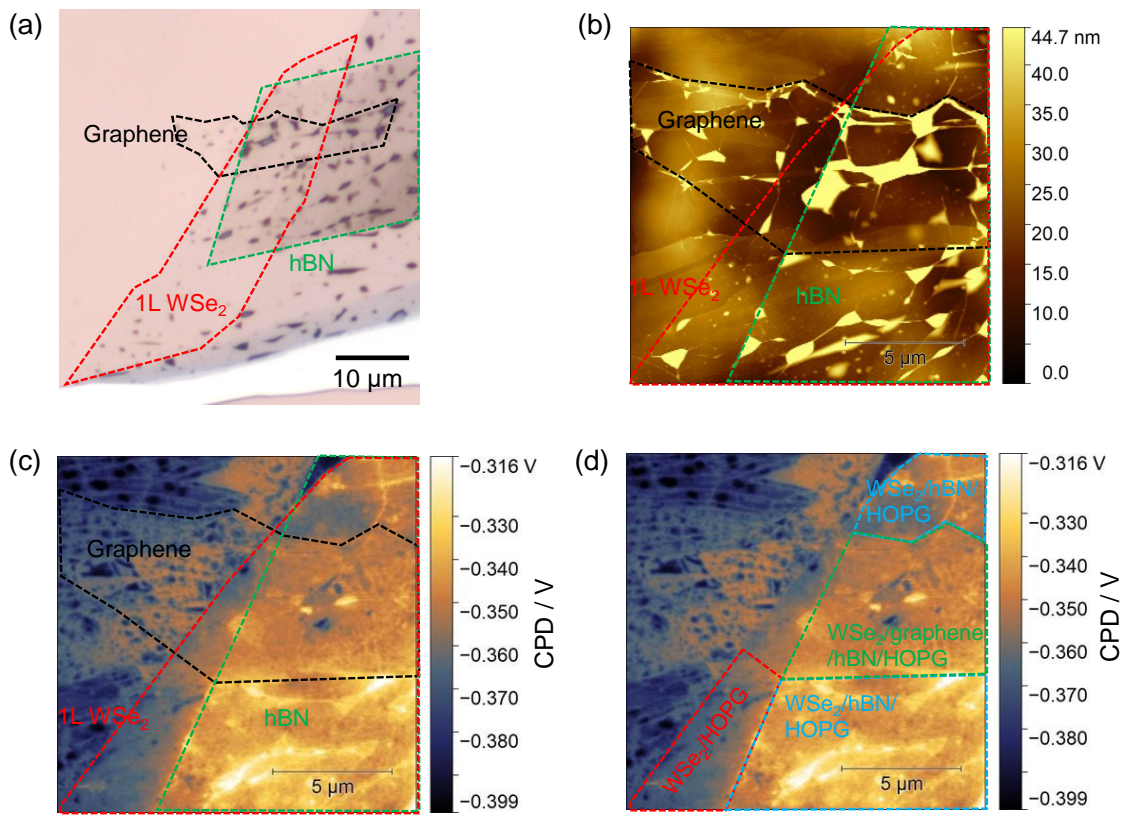

Figure 5S: (a) optical microscope image, (b) AFM height image, and (c-d) KPFM image of $WSe_2$/graphene/hBN/HOPG hetero-stack.

4

# 4  Work function determination of HOPG

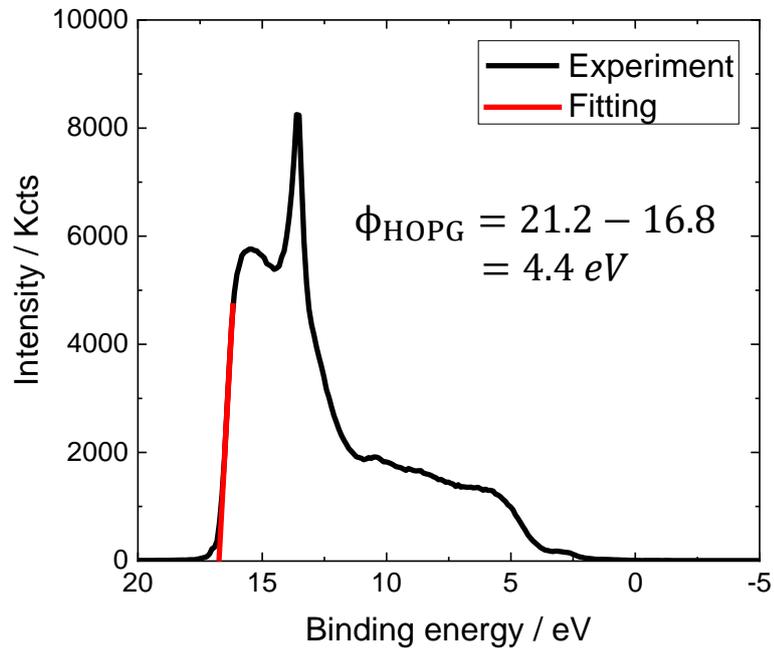

Figure 6S: UPS spectra of HOPG.

We use ultraviolet photoelectron spectroscopy (UPS) to determine the absolute work function of HOPG [8]. The He-I light source has an energy of 21.2 eV and the secondary electron cutoff (SEC) is 16.8 eV. The work function of HOPG is 4.4 eV.



\end{document}